# Effect of SOI substrate on silicon nitride resistance switching using MIS structure

A. Mavropoulis [a], N. Vasileiadis [a,b], C. Theodorou [c], L. Sygellou [d], P. Normand [a], G. Ch. Sirakoulis [b], P. Dimitrakis [a,*]

[a] *Institute of Nanoscience and Nanotechnology, NCSR "Demokritos", Ag. Paraskevi 15341, Greece*
[b] *Department of Electrical and Computer Engineering, Democritus University of Thrace, Xanthi 67100, Greece*
[c] *Univ. Grenoble Alpes, Univ. Savoie Mont Blanc, CNRS, Grenoble INP, IMEP-LAHC, 38000 Grenoble, France*
[d] *Institute of Chemical Engineering Sciences, FORTH/ ICE-HT, Patras 26504, Greece*

ABSTRACT

Several resistive memory technologies (RRAMs) are prominent, but few are fulfilling the requirements for CMOS integration and meet the commercialization standards. In this work, the fabrication and electrical character-ization of a fully compatible CMOS process on SOI substrate of 1R silicon SiN-based resistance switching (RS) MIS devices is presented. The RS characteristics are compared with the same devices previously fabricated on bulk silicon.

## 1. Introduction

Resistive memories (RRAMs) are one of the most promising nonvolatile memory alternatives. RRAM cells are perfect for implementation in crossbar architecture [1] that is able to achieve the smallest possible memory cell [2]. Such an implementation is rather appealing for in-memory and neuromorphic computing accelerators [3] especially for low power mobile IoT edge computing hardware technology [4].

Metal oxides (OxRAM) and phase-change materials (PCRAM) are the most studied materials from a technological standpoint, nitrides however, especially $SiN_x$ were found to exhibit competitive resistance switching properties and attractive SiN-based RRAM devices have been recently demonstrated [5]. Moreover, $SiN_x$ has been also used for neuromorphic and in-memory computing [6]. In the majority of the publications, the RS $SiN_x$ structures were MIS, meaning that the bottom electrode was heavily-doped Si. Silicon nitride is more resistant to oxygen and metal atoms diffusion [7]. RS mechanism of $SiN_x$ has been investigated under different metals (active or top electrode) [8] and doping of Si substrate (inert or bottom electrode) [9,10]. In MIS RS structures, the current overshoot is strongly affected by the dielectric and parasitic capacitance [11]. Random telegraph noise measurements at SET in SiN 1R cells revealed that true random number generator hardware can be also realized [12,13]. The scope of this work is to benchmark the use of thin SOI film as bottom electrode compared to bulk Si substrates in single MIS RS cells (1R) utilizing DC and AC measurements.

## 2. Device fabrication and measurement setup

MIS 1R cells are fabricated on (i) an Si wafer and (ii) an SOI wafer having 200 nm BOX and 100 nm silicon film, both doped with Phosphorus by the same ion implantation conditions targeting $\rho < 0.005$ Ω·cm (Fig. 1(a)). A 7 nm $SiN_x$ layer was deposited by LPCVD at 810 °C, using $NH_4$ and $SiCl_2H_2$ as gas precursors. XPS analysis revealed a molar fraction of $x = 1.27$, very close to the stoichiometric value of 1.33 (Fig. 1 (b)). 30 nm thick Cu was sputtered as top electrode, covered by Pt (30 nm) to prevent oxidation. A Tektronix 4200 SCS and a HP4284A precision LCR meter were used to perform DC *I-V* measurements and impedance measurements, respectively. For pulsed RS switching measurements, a custom setup was built (more detail in [7]). The back electrode (BE) contact was realized by Al on the backside of the chip die and in the frontside for bulk Si cells and SOI cells, respectively. BE was always grounded during measurements. Finally, current noise measurements were obtained in both frequency and time domain, using the Synergie Concept NOISYS7 system, for a bandwidth of 10 Hz to 10 kHz.

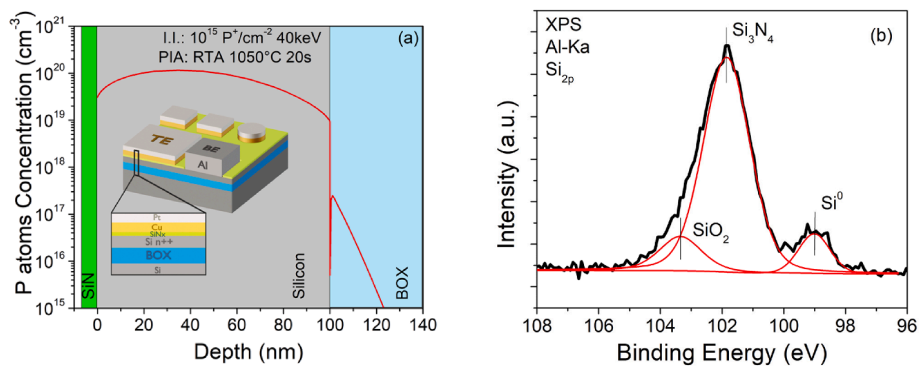

**Fig. 1.** (a) Synopsys Sentaurus TCAD simulation of the doping concentration in SOI. Inset, represents a schematic of the tested structures. (b) XPS spectra of the deposited SiN film by LPCVD.

## 3. Results and discussion

### 3.1. Current-Voltage sweep measurements

Multiple sequential SET/RESET voltage sweep cycles at different current compliance levels ($I_{cc}$) for the bulk and SOI MIS 1R cells are shown in Fig. 2, indicating their potential for resistance switching at multi-levels. A clear bipolar operation is observed for both bulk and SOI 1R cells with almost symmetrical $V_{SET}/V_{RESET}$ in the range [+3V,+3.5 V]/[-3.5 V,-3V]. Nevertheless, the resistance window is significantly lower for SOI cells, which is related to their lower HRS value compared to cells on bulk Si. Evidently, the current overshoot after SET is more pronounced in the case of bulk Si cells, meaning that SOI devices exhibited self-compliance characteristics [9]. This is mainly attributed to the parasitic capacitance due to prober's chuck capacitance and the bulk Si substrate resistance [14].

### 3.2. Multi-Level resistance switching for analog operation

Fig. 3(a) shows the resistance window (~5.3MΩ) for SOI MIS 1R cell at 0.3 V reading voltage. This can be achieved through an incremental-step-pulse-programming (ISPP) tuning protocol beginning from + 5 V (SET)/− 5V (RESET)/1 μs pulse and incrementally increasing/decreasing by + 0.1 V – 0.1 V/1 μs, respectively, until a maximum of + 7 V/− 7V occurs or a very low LRS state appears (indicating a first threshold of the material degradation) and resistance tuning procedure stops. A detailed description of the tuning protocol is described in [7]. Furthermore, using a modified version of the tuning algorithm described in [12], five stable resistance levels were found, as shown in Fig. 3(b). These intermediate resistance levels are in the range of those achieved by voltage sweeps applying different $I_{cc}$ (Fig. 2(b)), suggesting a promising analog memory behavior of our RRAM device.

### 3.3. Impedance measurements

Impedance measurements in the range 100 Hz – 1 MHz were performed for bulk and SOI cells at fresh (without any previous measurement), HR and LR states at 0.1 V. The devices were SET/RESET following voltage sweeps (see Fig. 2) with $I_{CC} = 100μA$. For both substrates, as shown in Fig. 4(a), the Nyquist plots at fresh and HR states of the cells are similar denoting that during RESET the RS material is fully recovered to its pristine state. However, the resistance at low frequencies representing the serial resistance between the two substrates is significantly different: 20 (70) Ω and 10 (20) kΩ for fresh (reset) states of bulk Si and SOI. Furthermore, the modelling of the Nyquist plots suggest that the mechanism should be described by a modified Warburg impedance element ($Z_W$) or a constant phase element ($Z_{CPE}$) which is attributed to a diffusion process [7]. Additionally, the Nyquist plots for LRS show that bulk Si MIS 1R cells are SET at lower resistance compared to SOI cells due to the enhanced current overshoot and the series resistance. LRS Nyquist graphs are semi-circles and thus can be modelled successfully by an equivalent circuit with a resistance in parallel to a capacitance, $R_p || C_p$ and a serial resistance $R_s$, as described in the inset of Fig. 4(b). Physically, $R_p$ and $C_p$ correspond to the resistance of the conductive paths formed during SET and the capacitance of the remaining insulating (non-switched) material region, respectively.

### 3.4. Noise measurements

Current fluctuations in time domain were measured at LRS for devices on both types of substrates. Fig. 5(a) shows two examples of such timeseries, where insets present the normalized occurrence histogram of current values. Fig. 5(b) demonstrates the power spectral density $S_I$ ($A^2$/Hz) of these timeseries normalized by the square of the average current, $I^2$, allowing for the comparison with noise density generated when the memristors are set at different resistance levels. The origin of the noise signal (current fluctuations) is due to the carrier exchange between the conductive filament and the surrounding $SiN_x$ traps [12,15–17]. Obviously, this is equivalent to the noise signal generated at the gate dielectric/channel of a MISFET and specifically to a gate-all-around nanowire MISFET. In this context, the approximation that $S_I$ is proportional to $I^2$ is considered valid [17,18], similar to MISFET theory where $S_{Id}$ is proportional to $I_d^2$. A careful observation of PSD plots reveals that PSD resulted from frequency domain measurements, $S_F$, is always higher the PSD obtained by time domain measurements, $S_T$. This is mainly attributed to the long-integration time used in frequency domain measurements in order to obtain smoother spectra. During this elongated measurement time the local temperature increases and thus the carrier exchange between the electrons flowing in the filament and the surrounding traps increases. Additionally, the density of thermal noise increases too. Moreover, $S_{F,T}$(SOI) is always higher than $S_{F,T}$(bulk-Si) because the read voltage in SOI devices was 0.2 V compared to 0.1 V bulk-Si devices. This was mandatory due to the limitation of the current amplifier in the lowest input current.

Evidently, all PSD have the form of $1/f^\gamma$, which is commonly attributed to carrier trapping/de-trapping in uniformly distributed slow traps. The inset table presents the values of the exponent $\gamma$, which is found to be very close to 1 for frequencies below 4 kHz, denoting the presence of a uniform trap distribution. For frequencies higher than 4 kHz, the spectrum is attenuated due to the limited bandwidth of the noise amplifier, related to the low measured current. For $\gamma$-values higher enough than 2 should be attributed to the impact of current amplifier's internal noise due to the level of the measured current values. The comparison between SOI and bulk Si reveals that the values of $\gamma$ are very close, suggesting that the trap distribution remains unchanged by the introduction of the SOI substrate as well as that the measured noise signals are mainly attributed to the SiN$_x$ traps.

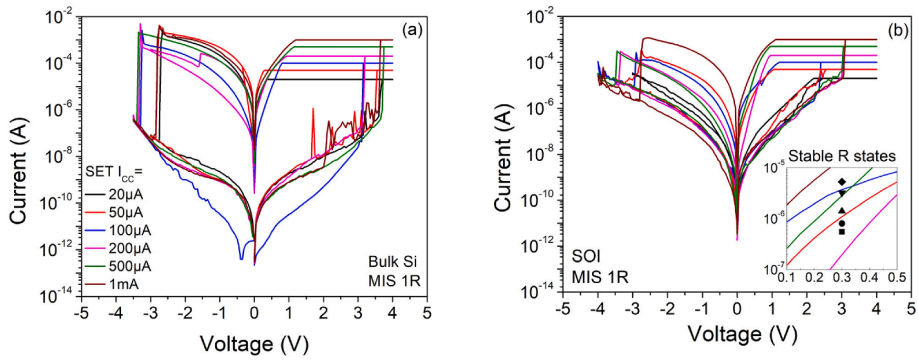

**Fig. 2.** *I-V* switching characteristics at different $I_{cc}$ for (a) bulk and (b) SOI MIS 1R cells. Inset plot indicates the stable resistive states shown in Fig. 3(b).

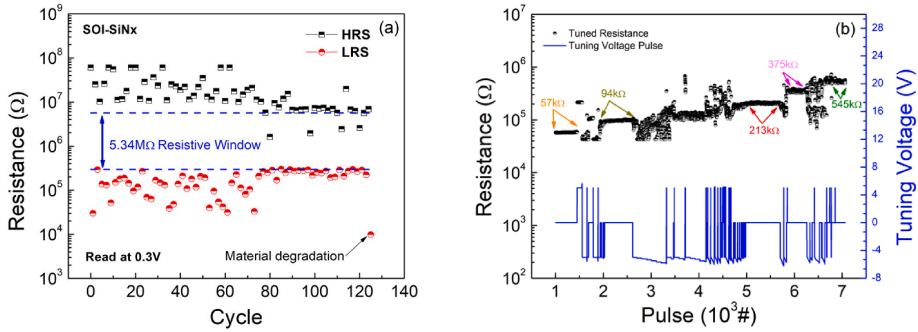

**Fig. 3.** (a) Endurance plot of an SOI MIS 1R cell using ISPP and (b) stable resistance levels obtained using protocol of [12].

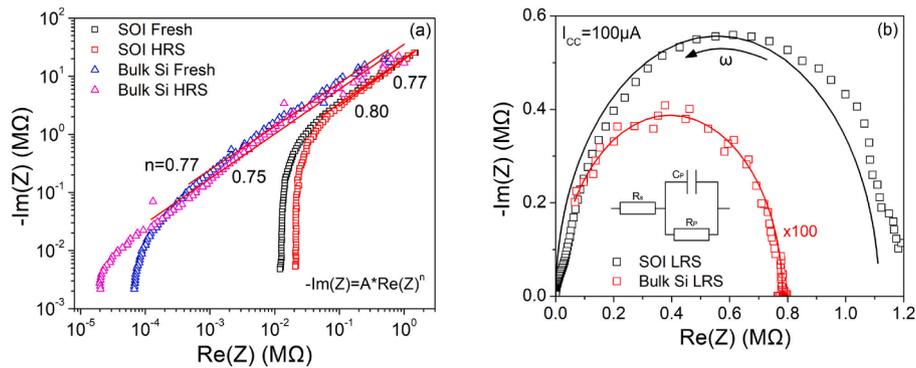

**Fig. 4.** Nyquist plots at (a) Fresh and HRS and (b) LRS, for bulk Si and SOI MIS 1R cells.

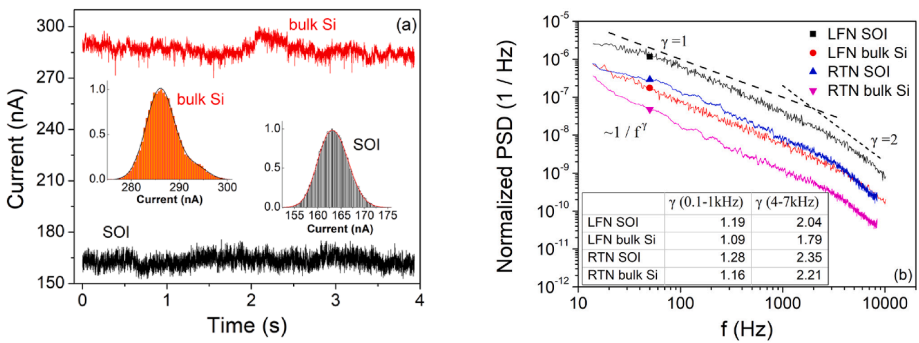

**Fig. 5.** (a) Noise current in time domain recorded with a sampling time 60 μs for $SiN_x$ memristors on SOI and bulk Si substrates measured at 0.2 V and 0.1 V respectively. Inset denotes the occurrence probability histogram. (b) Normalized PSD ($S_I/I^2$) as measured by noise analyzer and calculated from plot (a) for SOI and bulk Si substrates. For the sake of comparison, the PSD spectra have been smoothed by adjacent averaging with 100 measurements.

## 4. Conclusions

Silicon nitride memristors were fabricated on bulk Si and SOI heavily doped substrates acting as bottom electrode having Cu as top electrode. The series resistance of the bottom electrode was found to be higher on SOI compared to Si wafer, while the SOI substrate devices exhibited self-compliance characteristics as revealed by *I-V* voltage sweeps and AC impedance measurements. LFN spectral analysis indicated that there is no additional group of characteristic traps related to the SOI substrate.

**Declaration of Competing Interest**

The authors declare the following financial interests/personal relationships which may be considered as potential competing interests: This work was supported in part by the research projects "3D-TOPOS" (MIS 5131411) Operational Programme NSRF 2014-2020, General Secretariat of Research and Innovation (GSRI), Ministry of Development and "LIMA-chip" (Proj.No. 2748) which are funded by the Hellenic Foundation of Research and Innovation (HFRI).

**Acknowledgements**

This work was supported in part by the research projects "3D-TOPOS" (MIS 5131411) and "LIMA-chip" (Proj.No. 2748) which are funded by the Operational Programme NSRF 2014-2020 and the Hellenic Foundation of Research and Innovation (HFRI) respectively.